\documentclass[runningheads]{svmult}

\usepackage{makeidx}   % allows index generation
\usepackage{graphicx}  % standard LaTeX graphics tool
\usepackage{subeqnar}  % subnumbers individual equations
\usepackage{multicol}  % used for the two-column index
\usepackage{physprbb}  % modified textarea for proceedings,
\makeindex             % used for the subject index
\begin{document}
\title*{The Slow Merger of Massive Stars}
\toctitle{The Slow Merger of Massive Stars}
% allows explicit linebreak for the table of content
%
%
\titlerunning{The Slow Merger of Massive Stars}
% allows abbreviation of title, if the full title is too long
% to fit in the running head
%
\author{Natalia Ivanova\inst{1}
\and Philipp Podsiadlowski\inst{2}}
\authorrunning{Natalia Ivanova \& Philipp Podsiadlowski}
% if there are more than two authors,
% please abbreviate author list for running head
%
%
\institute{ Northwestern University, Evanston IL 60209, USA
\and  University of Oxford, Department of Astrophysics, Oxford OX1 3RH, UK}

\maketitle              % typesets the title of the contribution

\begin{abstract}
We study the complete merger of two massive stars inside a common
envelope and the subsequent evolution of the merger product, a rapidly
rotating massive supergiant.  Three qualitatively different types of
mergers have been identified and investigated in detail, and the
post-merger evolution has been followed to the immediate presupernova
stage.  The ``quiet merger'' case does not lead to significant changes
in composition, and the star remains a red supergiant.  In the case of
a ``moderate merger'', the star may become a blue supergiant and end
its evolution as a blue supergiant, depending on the core to total
mass ratio (as may be appropriate for the progenitor of SN~1987A).  In
the case of the most effective ``explosive merger'', the merger product
stays a red giant.  In last two cases, the He abundance in the
envelope is increased drastically, but significant s-processing is
mainly expected in the \mbox{``explosive merger'' case.}
\end{abstract}

\section{Introduction}
It is evident that the internal structure of the progenitor of a
core-collapse\index{core-collapse supernova} 
supernova (SN) is one of the dominant factors that
determines the characteristics of the supernova explosion, such as the
light-curve and the abundances produced in the supernova (see
e.g.\ \cite{Muller}).  It has also been shown that the distribution of
the angular velocity can produce a strong asymmetry in the
nucleosynthesis during the SN explosion and in its ejecta
\cite{Fryer}. This makes it necessary to follow the detailed evolution
of the abundances and the rotation profile at all stages of the
evolution of a massive star before it explodes as a core-collapse SN.
Observationally, it is well established that $\sim 40\,$\% of all
massive stars are members of binary\index{binary} systems with orbital periods
shorter than 1 year and that at least 25\,\% of these will start to interact
by Roche-lobe overflow (RLOF) during the advanced stages of the
primary's evolution \cite{Garmany,PJH}.
% Theoretical constrains, given by the binary population synthesis, 
% confirm that majority of the massive stars are formed as 
% binary components with orbital period between 1 day and 10 yr and
% most of them interact \cite{Vanbeveren}. 
This implies that a significant fraction of all core-collapse
supernova progenitors will have been affected by a previous binary
interaction, where one of the most important interactions is the
spiral-in of the two binary components inside a common envelope (CE)
\cite{PJH}.  The final result of the spiral-in depends on how much of
the released orbital energy has been converted into driving the
expansion of the envelope relative to its binding energy.  Here we
study the situation where the deposited energy is not sufficient to
eject the common envelope\index{common envelope}.  This leads to the
complete merger\index{binary merger} of the secondary with the core of
the primary, forming a rapidly rotating single star in the process.
Since the timescale of the merger is much longer than the dynamical
timescale of the CE, it cannot be treated with a purely hydrodynamical
code.  Mass transfer in the merging phase changes not only the
chemical composition profile and the angular velocity distribution,
but can also cause the erosion of part of the core, changing the
core/envelope ratio.  The evolutionary path of the merger product may
also differ from that of a normal single star, e.g. by making a late
blue loop more probable. Since the resulting supernovae may also differ
from those with single-star progenitors, these merged objects are
likely to be responsible for some of the variety observed among Type
II supernovae.

In this paper we present the results of the modelling of the complete slow
merger of a massive binary within a common envelope. In Section 2 we
briefly describe the assumptions that were used to model mergers, and
in Section 3 we present some of the main results.

\section{Method and Initial Models}
We used a standard Henyey-type stellar evolution code \cite{KWH},
updated recently \cite{PRP}.  The nuclear reactions rates were taken
from Thielemann's library REACLIB \cite{TTA} and updated as in
\cite{Cannon}.  In the code OPAL opacities \cite{RI} are used,
supplemented with contributions from atomic, molecular and grain
absorption in the low temperature regime \cite{AF}.

To model the merger we implemented a number of modifications to the
single stellar evolution code.  These modifications were made mainly
to treat the presence of the secondary inside the primary's envelope,
including the mass transfer from the secondary to the core and the
associated nucleosynthesis and mixing.  A more detailed description of
the modifications in the code can be found in \cite{IP}. We determine
how deep the hydrogen-rich stream penetrates into the core of the
primary using the prescription developed in \cite{IPS}.

We considered binaries consisting of a $18-22\,M_\odot$ primary and a
$1-5\,M_\odot$ secondary.  At the start of the spiral-in, the primary
had already completed core helium burning.  The chemical composition
was taken as typical for young stars in the LMC ($X=0.71$ and
$Z=0.01$).  We adopted the Schwarzschild criterion for convection and
took a mixing-length parameter $\alpha = 2$ and a
convective-overshooting parameter equal to 25\,\% of a pressure scale
height. These parameters were chosen since they are most appropriate
for merger models of the progenitor of SN 1987A\index{SN 1987A}
(see \cite{PODSI}).

\section{Results}

The qualitative behaviour of the merger and the temporal evolution of
the structure of the primary within the secondary's orbit depend on
the interaction of the hydrogen-rich material with the surrounding
ambient matter. In particular, it depends on how deep and how fast the
hydrogen-rich material penetrates into the primary's core and its
placement with respect to the hot and/or convective zones within the
secondary's orbit.  According to our prescription for determining the
stream penetration depth, the most effective penetration should occur
near the end of the merger phase, when the mass-loss rate is high and
the exposed material from the secondary has low entropy. However, at
this time the structure of the primary has changed, and a dense
hydrogen-enriched region may already have been built up around the
helium core, preventing the stream from penetrating deeper.  All these
effects combined create a very non-linear picture of the primary's
response to mass transfer. In general, as a result of the merger, the
primary's core expands, and the central temperature of the core drops
(the degree of core cooling depends on the core expansion, i.e.\
the merger efficiency). This increases the total
evolutionary time before core carbon ignition.

Based on our systematic study, we can distinguish three qualitatively
different types of merger, where the classification of the mergers
can be well explained by considering the temporal evolution of the
convective zones during the merger.\\

\noindent {\bf $\bullet$  The quiet merger. }
All of the He affected by the penetrating stream is mixed with the
outer envelope during the merger by convection, but since most of the
He shell is not disturbed, there is only a moderate change in the
surface abundances (He at the surface increases only by $4-8\,\%$).
The merger product remains a red supergiant, although the progenitor
might be significantly spun up compared to a single supergiant evolved in
isolation. \\
This type of merger may happen in systems where the primary is close
to carbon ignition at the start of the merger. This implies high
pressure and temperature gradients and a correspondingly high entropy
dissipation coefficient \cite{IPS}. In addition, the secondary has to
have a mass larger than $\sim 2\,M_\odot$ (due to the larger entropy).\\

\noindent {\bf $\bullet$  The moderate merger. }
\begin{figure}[top]
\includegraphics[scale=0.40]{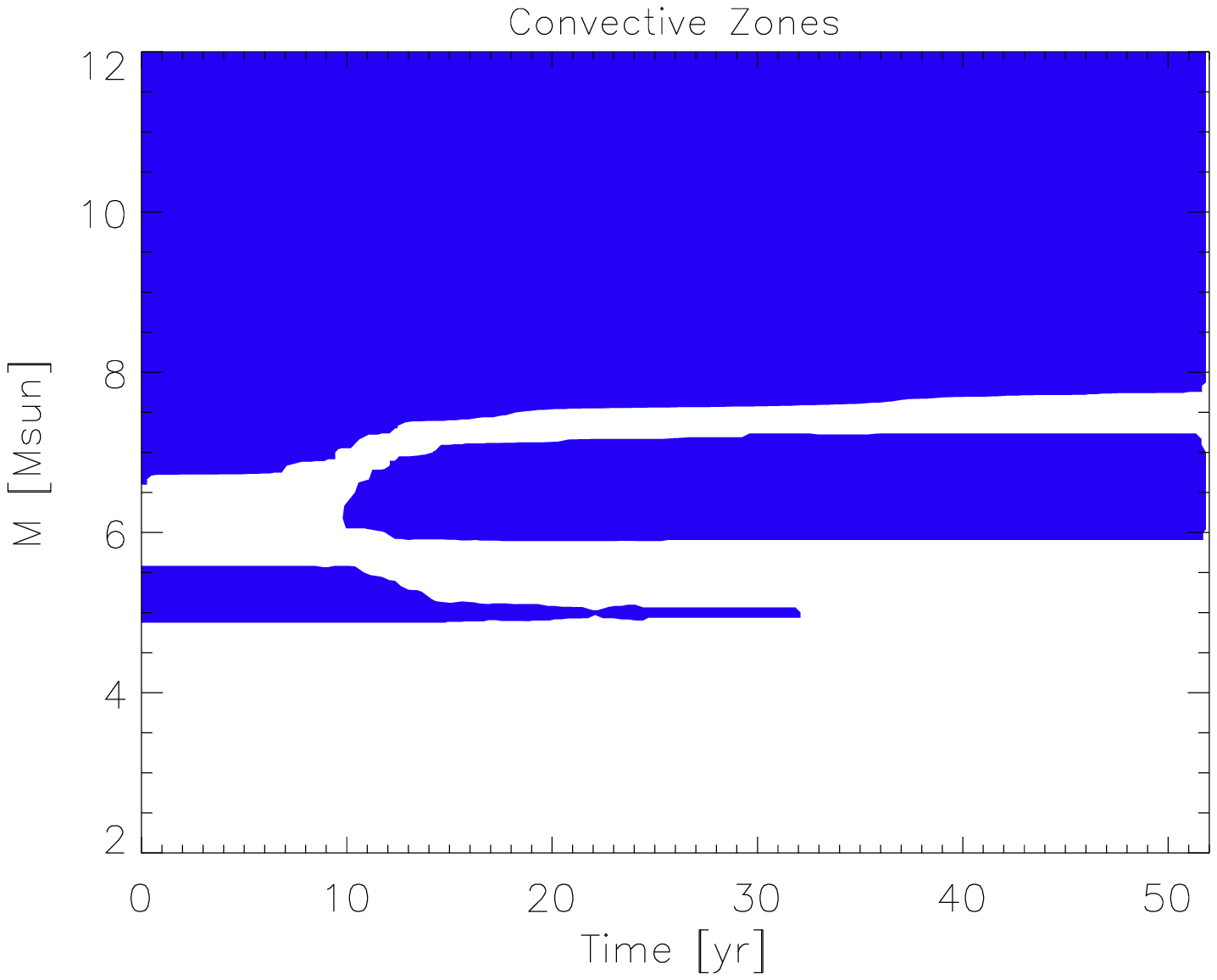}
\includegraphics[scale=0.40]{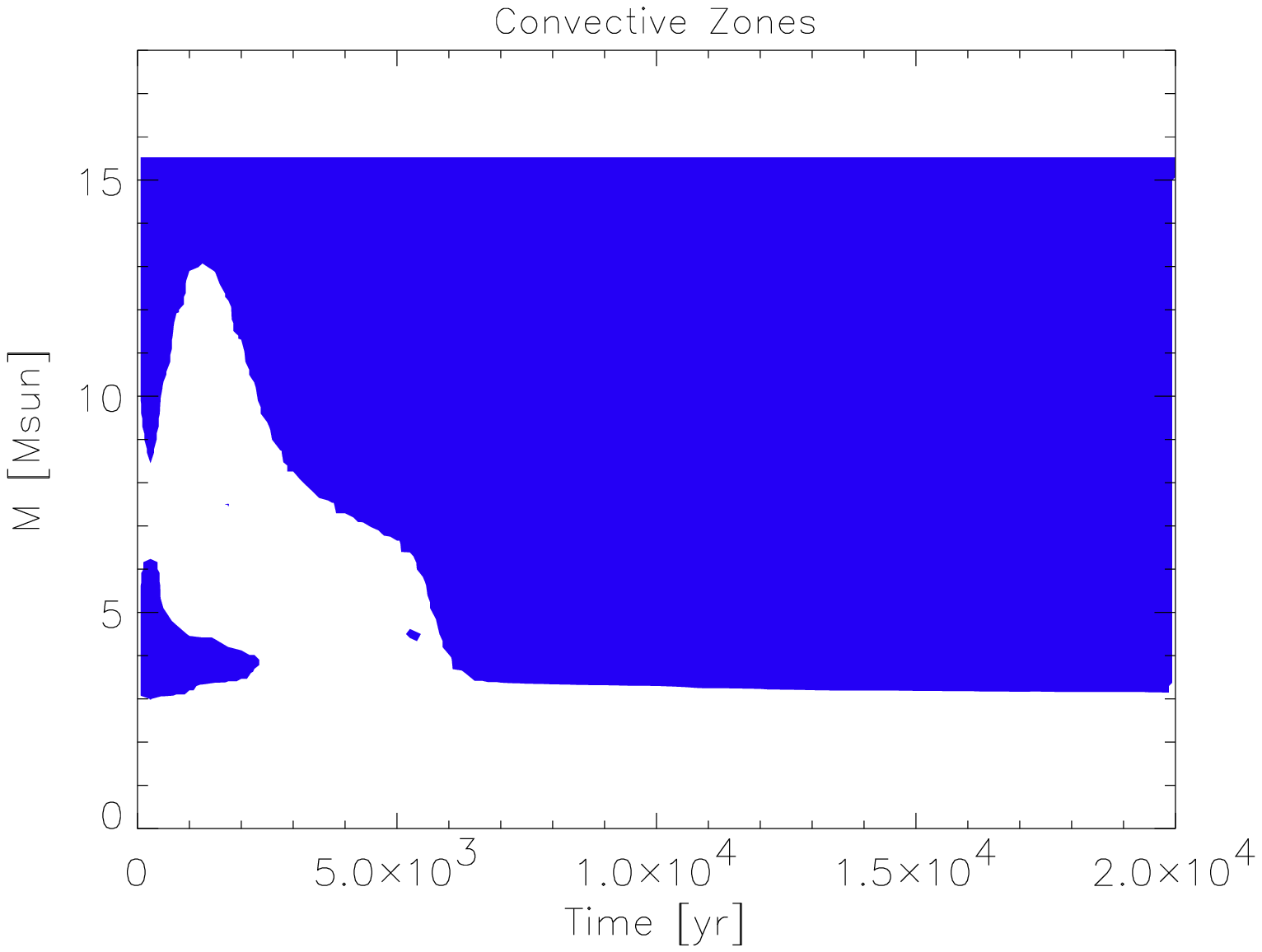}
\vspace{-30pt}
\caption{
The moderate merger: the evolution 
of convective zones (by mass)
during the merger (left panel) and after the merger until core carbon
ignition (right panel).}
\end{figure}
Here the He core expands significantly, but an extensive He-rich shell
remains.  During the penetration, the stream creates a hydrogen-rich
zone around the core. This zone becomes convective at some point and
suppresses the bottom helium convective zone (see Fig.~1). During the
merger, the primary core expands more drastically than in the case of
the quiet merger.  The merger product appears as a red supergiant,
rotating rapidly and contracting immediately after the merger.
Significant rotationally enhanced mass loss in the equatorial
direction during the contraction phase is expected.  Depending on the
core/envelope ratio, the merger product may then perform a blue loop
(if the hydrogen shell source becomes temporarily dominant) or
continues its evolution as a red supergiant.  In the first case, some
dredge-up of helium takes place during the merger, and it can be
expected that rotationally induced mixing will cause further
significant enhancement of the surface helium abundance.  In the
second case, a delayed dredge-up phase (a few thousand years after the
merger, see Fig.~1) ) takes place, resulting in a large overabundance
of He in the envelope ($20- 80\,\%$).\\ 
At the start of the merger, the primary can be either already close to
core carbon ignition and has a companion of $\sim 1\,M_\odot$, or
the primary is at the start of He shell burning and has a companion 
of $2\,M_\odot$ or larger. A moderate merger is most appropriate
for the progenitor of SN 1987A\index{SN 1987A}.\\

\noindent {\bf $\bullet$ The explosive merger. }
During the stream-core interaction, the hydrogen-rich material
accumulates between two initially convective zones and slowly creates
an intermediate zone, which at some point connects to the He burning
shell. This leads to a dramatic nuclear flash, resulting in a drastic
expansion of the He shell and complete mixing. The duration of this He
shell explosion is about 0.25 year.  In the case of a strong He shell
explosion, even the carbon core may be disturbed. In all cases, there
is an immediate significant increase of helium in the envelope, often
accompanied by an increase in the carbon abundance.  The stripped-off
naked carbon core connects with the hydrogen-rich convective envelope
and provides the site for efficient s-processing.  The merger product
continues its evolution as a red supergiant.\\
This type of the merger can take place in binaries with 
low-mass secondaries, not very close to core carbon ignition.
The anomalous carbon star V Hydrae may provide an example for this
merger channel.

%\vspace{-12pt}

\end{document}